\documentclass[11pt]{article}

\usepackage[preprint]{acl}

\usepackage{times}
\usepackage{latexsym}
\usepackage{amssymb}
\usepackage{booktabs}
\usepackage{tabularx}
\usepackage{multirow}

\usepackage[T1]{fontenc}

\usepackage[utf8]{inputenc}

\usepackage{microtype}

\usepackage{inconsolata}

\usepackage{graphicx}

\title{Autonomy Is Absent, Not Eroded: An Observational Study of User Autonomy in Multi-Turn LLM Conversations}

\author{
Chaoran Chen \\
{\small University of Notre Dame} \\
\And
Isabel Barberá \\
{\small Independent Researcher} \\
\And
Yuhuai Wei\thanks{The authors are ordered alphabetically by the initials of their first names.} \\
{\small Independent Researcher}
}

\begin{document}
\maketitle

\begin{abstract}
Does prolonged interaction with Large Language Models reduce users' observable autonomy during conversation? We test this \emph{degradation hypothesis} by scoring user autonomy across 2,877 multi-turn conversations (46,952 turns) from WildChat and validating on 500 conversations from LMSYS-Chat-1M. Our core finding is a null result: population-level autonomy remains stable throughout conversations, only 4.7\% of conversations show meaningful decline, and an apparent decrease at later turn positions is explained by survivorship bias. Instead, the dominant pattern is not erosion but \emph{absence}: 80.8\% of conversations exhibit zero or low autonomy throughout. When autonomy engagement occurs, it is self-reinforcing (persistence doubles from 40\% to 85\% over consecutive turns). These findings reframe the problem: the risk is not that LLMs progressively degrade users' observable autonomy, but that most users never engage in independent reasoning during the interaction.
\end{abstract}

\section{Introduction}

There is widespread concern that interacting with LLMs reduces users' autonomy---that the more turns a conversation continues, the more users surrender independent judgment to the model~\cite{10.1145/3706598.3713778,10.1016/j.chb.2024.108352}. This concern gives rise to what we refer to as the \emph{degradation hypothesis}: that users become progressively less autonomous as conversations continue. The hypothesis motivates safety work on dark patterns~\cite{kran2025darkbench,barbera2026dgamm}, sycophancy~\cite{10.1126/science.aec8352,sharma2024towards}, and persuasive manipulation~\cite{Matz2024,Salvi2025} in LLMs. Yet the hypothesis has not been tested observationally: do users actually become less autonomous over the course of real conversations? While this question does not address longer-term changes in autonomy across repeated LLM use, it is a necessary first step in understanding whether autonomy declines during individual interactions.

We conduct a large-scale observational study to find out. We score user autonomy across 2,877 multi-turn conversations (46,952 user turns) from WildChat~\cite{zhao2024wildchat} using an LLM judge calibrated against a three-dimensional rubric, and validate on 500 conversations from LMSYS-Chat-1M~\cite{zheng2024lmsyschatm}. Our central finding is a \textbf{null result for degradation}: observable autonomy remains stable across conversation progress, and an apparent decline at later turn positions is entirely attributable to survivorship bias. The actual dominant pattern is \emph{absence} rather than erosion: most users exhibit little or no observable autonomy throughout the interaction.

Specifically, we find: (1) 80.8\% of conversations show zero or low autonomy throughout, with 56.8\% of turns being cognitive offloading; (2) only 4.7\% of conversations show temporal decline; (3) assistant-side Directive Pressure has negligible effect on user autonomy (Cohen's $d = 0.04$); (4) autonomy is self-reinforcing---persistence doubles from 40\% to 85\% over consecutive autonomous turns; and (5) these patterns replicate on LMSYS-Chat-1M across diverse open-source models.

\section{Method}

\subsection{Dataset and Filtering}

We draw on the WildChat dataset~\cite{zhao2024wildchat} (529,428 conversations from real ChatGPT users). We filter for English conversations with $>$12 interaction rounds to enable within-conversation temporal analysis, then remove exam dumps, repetitive tool-use patterns, minimal-engagement conversations, and non-English residuals (full pipeline in Appendix~\ref{app:filtering}). The final dataset comprises 2,877 conversations with 46,952 scored user turns (mean 17.4 rounds, 95.6\% GPT-3.5-turbo).

\subsection{User Autonomy Scoring}
\label{sec:autonomy_scoring}

We develop a per-turn scoring rubric to operationalize observable user autonomy during interaction. The rubric distinguishes three turn types:

\paragraph{Mechanical (M).} The turn involves no decision space---providing input data, factual constraints, or well-defined instructions (e.g., ``Here's the error message: [paste]'').

\paragraph{Offloading (O).} The user delegates a cognitive task they could reasonably engage with themselves, without expressing their own reasoning (e.g., pasting an exam question for the AI to answer).

\paragraph{Applicable (A).} The user demonstrates independent judgment through reasoning, evaluation, planning, or choice (e.g., ``I think that's wrong because X'').

For Applicable turns, three dimensions are scored on a \{0, 0.5, 1\} scale:

\begin{itemize}
    \item \textbf{Critical Engagement (CE)}: Does the user evaluate the assistant's output through verification, questioning, contradiction, or independent reasoning?
    \item \textbf{Option Exploration (OE)}: Does the user actively engage with the solution space---acknowledging alternatives and seeking trade-offs?
    \item \textbf{Decision Ownership (DO)}: Does the user retain final decision authority, or delegate it entirely to the assistant?
\end{itemize}

The per-turn autonomy score is:
\[
A_t = 0.3 \times CE + 0.3 \times OE + 0.4 \times DO
\]
with Offloading turns scored as $A_t = 0$ and Mechanical turns excluded from aggregation. The conversation-level score is the mean of all non-Mechanical turns.

Throughout this paper, autonomy refers to observable autonomy during interaction rather than users' underlying cognitive processes.

\subsection{Assistant-Side Directive Pressure}

To test whether assistant behavior is associated with user autonomy, we separately score assistant turns for Directive Pressure (DP) on a \{0, 0.5, 1\} scale:

\begin{itemize}
    \item \textbf{DP = 0} (Neutral): Presents multiple options with trade-offs or delivers factual content without prescribing action.
    \item \textbf{DP = 0.5} (Soft recommendation): Gives a recommendation but acknowledges alternatives.
    \item \textbf{DP = 1} (Strong prescription): Prescribes a single path with no mention of alternatives.
\end{itemize}

DP 
is scored only when the assistant's response involves advice or recommendations where alternatives could exist; responses to Mechanical user turns receive DP = 0 by default.

\subsection{LLM-as-Judge Implementation}

Both rubrics are applied using Claude Opus 4.6 as an LLM judge. The rubric is designed to evaluate observable behaviour expressed in the conversation rather than infer users' internal reasoning. Accordingly, the LLM judge evaluates only evidence expressed in the dialogue. The judge receives the full conversation context and scores each user turn sequentially, outputting structured JSON with classification, dimension scores, and brief rationale. A total of 2,877 conversations were scored for user autonomy (46,952 turns) and 1,418 conversations were scored for assistant-side Directive Pressure.

\subsection{Temporal Analysis}

To detect within-conversation autonomy change, we compute:
\[
\Delta_A = \bar{A}_{second\ half} - \bar{A}_{first\ half}
\]
where halves are defined over non-Mechanical turns. Conversations with $\Delta_A \leq -0.15$ are classified as \emph{degrading}, those with $\Delta_A \geq 0.15$ as \emph{improving}, and the remainder as \emph{stable}.

\section{Results}

\subsection{Turn-Level Classification}

Table~\ref{tab:turn_classification} shows the distribution of turn types across all scored conversations.

\begin{table}[t]
\centering
\small
\caption{Turn-level classification results (N = 46,952).}
\label{tab:turn_classification}
\begin{tabular}{lrrl}
\toprule
\textbf{Classification} & \textbf{Count} & \textbf{\%} & \textbf{Description} \\
\midrule
Offloading (O) & 26,661 & 56.8 & Zero own reasoning \\
Mechanical (M) & 13,517 & 28.8 & No decision space \\
Applicable (A) & 6,774 & 14.4 & Independent judgment \\
\bottomrule
\end{tabular}
\end{table}

Over half of all user turns are pure cognitive delegation. Only 1 in 7 turns shows evidence of independent thinking. Among conversations with valid conversation-level autonomy scores ($N = 2{,}631$; excluding conversations without eligible non-Mechanical autonomy scores), the mean autonomy score is 0.102 (median 0.029), with 42.0\% scoring exactly zero and 80.8\% falling in the zero-or-low category (score $< 0.2$).

\subsection{Population-Level Trajectory}

\begin{figure}[tp]
    \centering
    \includegraphics[width=0.92\columnwidth]{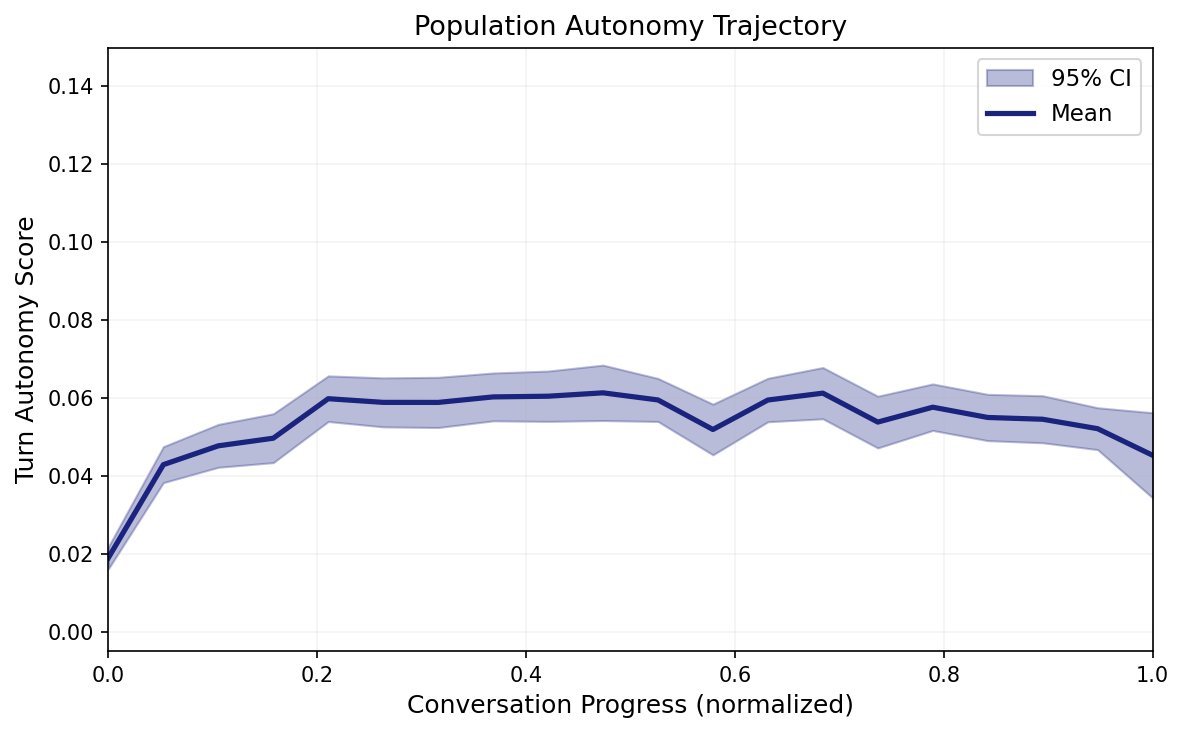}
    \caption{Population-level autonomy trajectory across normalized conversation progress. The curve is flat: no population-level degradation is observed. Shaded band shows 95\% CI.}
    \label{fig:population_trajectory}
\end{figure}

Figure~\ref{fig:population_trajectory} shows the mean turn autonomy score across normalized conversation progress. The trajectory is essentially flat, ranging from approximately 0.02 to 0.06, with no systematic decline. The population-level mean delta is $+0.011$---negligible and slightly positive.
Although mean autonomy scores are low in absolute terms, the absence of a temporal decline is consistent across our analyses.

\subsection{Temporal Dynamics}

Among conversations with sufficient non-Mechanical turns for temporal analysis, degradation is rare: 121 conversations show degradation ($\Delta_A \leq -0.15$), 177 show improvement ($\Delta_A \geq 0.15$), and 2,122 remain stable. When degradation does occur, it is typically abrupt rather than gradual: affected conversations show a mean first-half score of $\sim$0.5 collapsing to $\sim$0.1 in the second half.

\subsection{Survivorship Bias Explains an Apparent Decline}
\label{sec:survivorship}

\begin{figure*}[tp]
    \centering
    \includegraphics[width=0.95\linewidth]{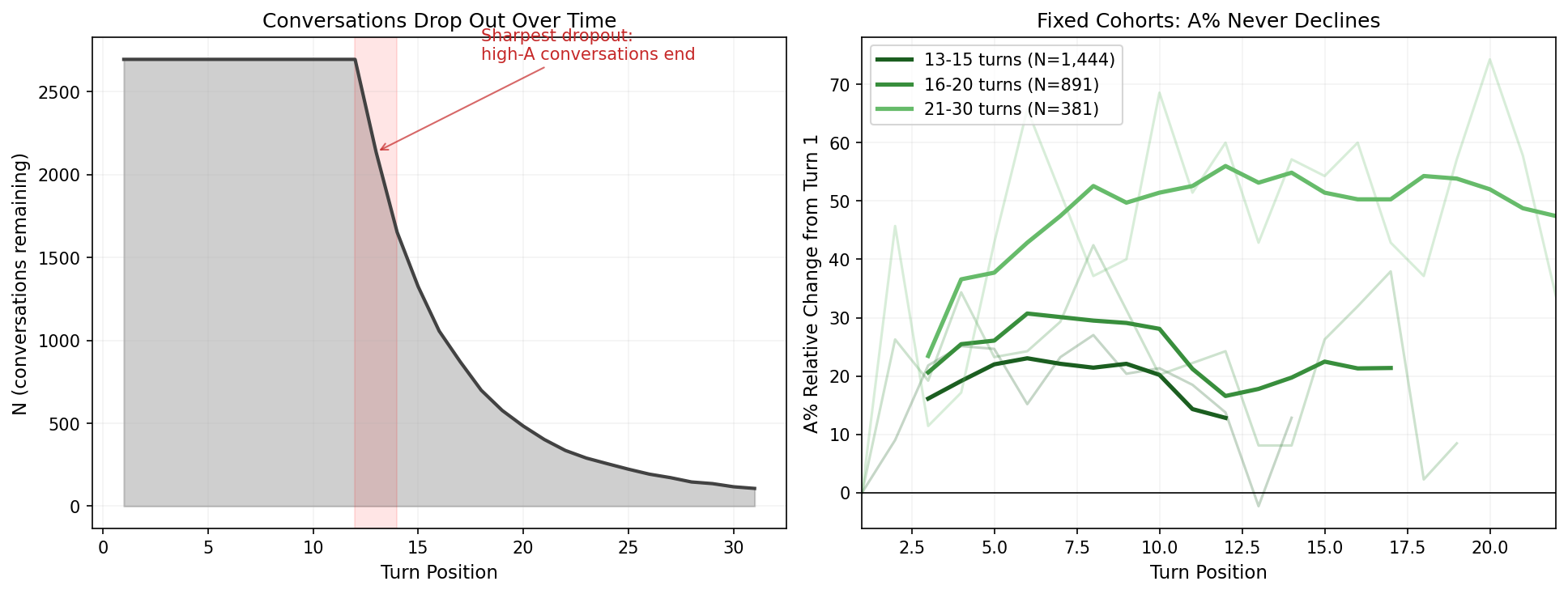}
    \caption{Left: Conversation dropout over turn positions. Right: Fixed-length cohort analysis---within cohorts of equal length, A\% never declines. The population-level ``decline'' is entirely compositional.}
    \label{fig:survivorship}
\end{figure*}

When tracking turn-type proportions by absolute position (Figure~\ref{fig:mode_decomposition} in Appendix~\ref{app:figures}), the Applicable proportion appears to collapse after turn 15. However, this is entirely attributable to differential dropout (Figure~\ref{fig:survivorship}): high-autonomy conversations end sooner, leaving low-autonomy conversations in later positions.

When we control for conversation length by analyzing fixed cohorts (13--15, 16--20, and 21--30 turns), A\% within each cohort never declines. The cross-sectional decline is a survivorship artifact, not a within-user effect.

\subsection{Assistant Directive Pressure: A Null Result}

Assistant-side Directive Pressure (DP) co-varies with A\% across turn positions (Figure~\ref{fig:dp_covariation} in Appendix~\ref{app:figures}), suggesting directive AI behavior might suppress autonomy. Three analyses reveal this is not causal:

\paragraph{Turn-level.} A-turn scores are virtually identical across DP levels: DP=0 ($\bar{s}=0.386$, $N=2{,}825$), DP=0.5 ($\bar{s}=0.391$, $N=3{,}387$), DP=1.0 ($\bar{s}=0.381$, $N=398$). Cohen's $d = 0.04$ (negligible). All sub-dimensions (CE, OE, DO) remain stable.

\paragraph{Within-conversation.} The correlation between $\Delta$DP and $\Delta$Autonomy across conversation halves is $r \approx 0$. Conversation-level correlation between mean DP and A\% is $r = 0.06$.

Full visualizations in Appendix~\ref{app:figures} (Figures~\ref{fig:null_effect},~\ref{fig:dp_correlation}).

\subsection{Autonomy Is Self-Reinforcing}

\begin{figure}[tp]
    \centering
    \includegraphics[width=0.92\columnwidth]{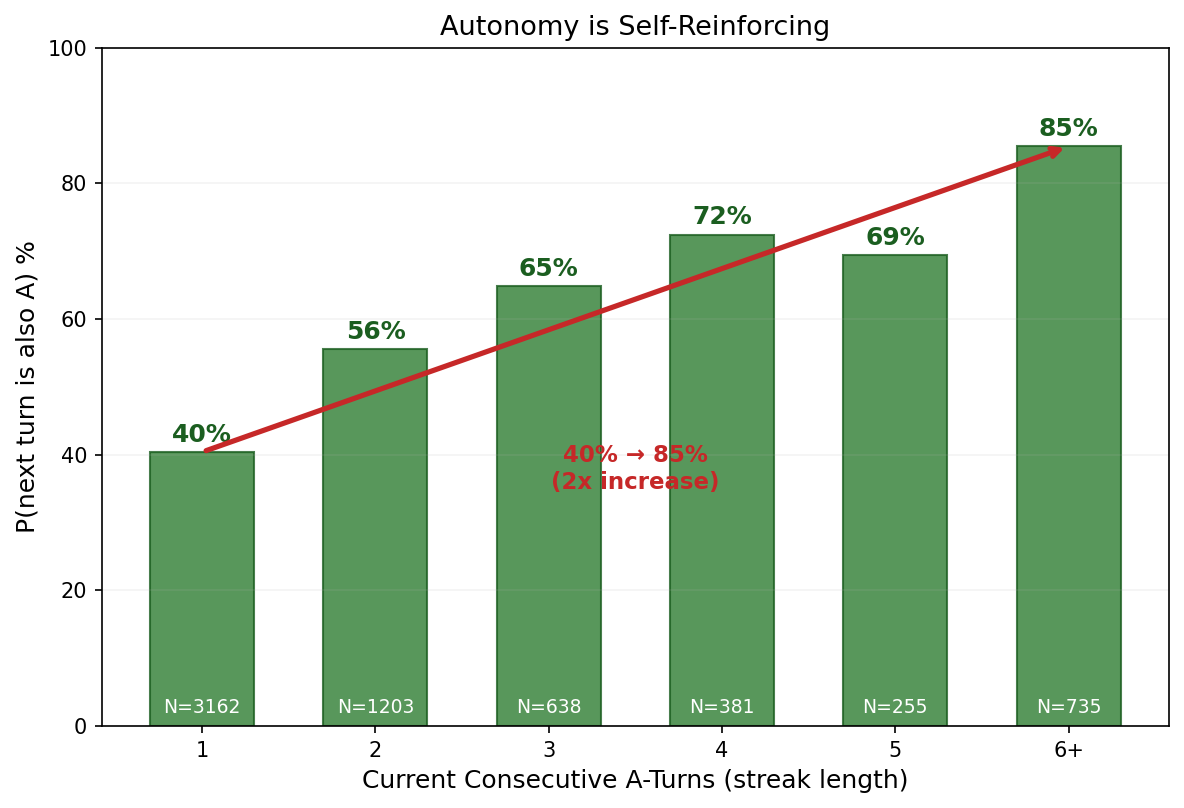}
    \caption{Autonomy is self-reinforcing. The probability of sustaining autonomous engagement doubles from 40\% to 85\% over consecutive A-turns.}
    \label{fig:streak}
\end{figure}

The one robust dynamic we observe is positive momentum: autonomy is self-reinforcing (Figure~\ref{fig:streak}). The probability that the next turn will also be Applicable given a current streak of consecutive A-turns increases from 40\% (streak = 1) to 85\% (streak $\geq$ 6)---a doubling that holds with large sample sizes at every point ($N$ ranges from 255 to 3,162).

\subsection{Within-Conversation Quality Stability}

A complementary analysis confirms that when users do exercise autonomy, the \emph{quality} of their autonomous engagement does not degrade (Figure~\ref{fig:stability} in Appendix~\ref{app:figures}). In paired early-vs-late comparisons within fixed-length conversations (13--15 turns), mean A-turn scores are virtually identical between turns 0--4 and turns 10+. The sub-dimension analysis reveals that Critical Engagement rises slightly over conversation progress (+0.11), while Option Exploration and Decision Ownership remain flat.

\subsection{Cross-Dataset Validation: LMSYS-Chat-1M}
\label{sec:crossval}

To test whether our findings are specific to WildChat or reflect broader patterns in human-LLM interaction, we replicate the core analyses on a stratified sample of 500 conversations from LMSYS-Chat-1M---a separate dataset covering diverse open-source models (predominantly Vicuna-13B, plus Koala, Alpaca, Claude-1, GPT-4, and others). We apply the same filtering pipeline and LLM judge scoring. Table~\ref{tab:crossval} summarizes the comparison.

\begin{table}[t]
\centering
\caption{Cross-dataset validation: WildChat vs.\ LMSYS-Chat-1M.}
\label{tab:crossval}
\resizebox{\columnwidth}{!}{%
\begin{tabular}{lrrl}
\toprule
\textbf{Finding} & \textbf{WildChat} & \textbf{LMSYS} & \textbf{Replicates?} \\
\midrule
Offloading \% & 56.8 & 37.6 & Partial \\
Applicable \% & 14.4 & 22.3 & Partial \\
Absence rate (score $<$0.2) & 80.8 & 65.7 & Yes \\
Mean conv score & 0.102 & 0.152 & Yes \\
Mean $\Delta$ & +0.011 & +0.011 & Yes \\
Degrading ($\Delta \leq -0.15$) & 4.7\% & 9.8\% & Yes \\
Streak=1 $\to$ P(A) & 40\% & 42\% & Yes \\
Streak=5 $\to$ P(A) & 81\% & 75\% & Yes \\
\bottomrule
\end{tabular}}
\end{table}

The core findings replicate: (1) absence remains the dominant pattern (65.7\% of LMSYS conversations show zero-or-low autonomy), (2) mean delta is near zero (+0.011), confirming no population-level degradation, (3) degradation is rare even at the 0.15 threshold (9.8\%), and (4) autonomy is self-reinforcing (streak persistence doubles from 42\% to 75\%). The main quantitative difference is that LMSYS shows somewhat higher Applicable rates (22.3\% vs. 14.4\%) and lower Offloading (37.6\% vs. 56.8\%), likely reflecting that LMSYS conversations are captured from a platform where users compare models---a context that may elicit more evaluative behavior. Despite these distributional differences, the trajectory-level conclusions hold.

\section{Discussion}

\subsection{Absence, Not Erosion: Implications for AI Safety}

The prevailing degradation hypothesis---that LLMs progressively erode users' autonomy during interaction---is not supported. We find no evidence of a systematic decline in observable autonomy over the course of conversations: population-level autonomy remains stable, only 4.7\% of conversations degrade, and an apparent decline at later turn positions is explained by survivorship bias. The actual risk is \emph{absence}: 80.8\% of conversations show zero or low autonomy throughout. Our findings instead suggest that the primary challenge is not progressive loss of observable autonomy during interaction, but the fact that many conversations begin—and remain—with little observable autonomous engagement.
This finding should not be interpreted as evidence that users do not reason during LLM interactions. Rather, they rarely express observable autonomous reasoning within the conversation itself.

This has concrete implications. First, for \textbf{safety evaluation}: benchmarks designed to detect trajectory-level degradation are measuring a rare phenomenon. The more productive evaluation target is whether systems can \emph{activate} autonomous engagement---since our streak analysis shows that once autonomy appears, it self-sustains. Second, for \textbf{AI design}: assistant directiveness (Cohen's $d = 0.04$) does not measurably affect user autonomy in naturalistic interaction. User-side factors---domain expertise, task type, individual disposition---appear to dominate. This does not exonerate deliberately manipulative designs, but suggests that routine directiveness is not the mechanism of concern. Third, for \textbf{scope}: our findings apply to within-session dynamics. Long-term dependency formation across sessions---the broader societal concern---may operate through habitual offloading reinforced over time, rather than acute within-conversation manipulation.

\subsection{A Methodological Caution: Survivorship Bias in Multi-Turn Corpora}

The survivorship bias we identify has implications beyond this study. Any cross-sectional analysis of multi-turn corpora (tracking a variable by turn position across all conversations) is susceptible to the same artifact: if conversations with property $X$ tend to be shorter, $X$ will appear to decline at later positions even if no individual conversation shows decline. Researchers studying engagement, toxicity, or satisfaction trajectories in conversational AI should control for conversation length before interpreting positional trends---as we demonstrate with fixed-cohort analysis.

\subsection{What Drives Autonomy?}

If assistant behavior does not predict user autonomy, what does? Our qualitative cases and quantitative patterns suggest three factors. First, \textbf{domain expertise}: the same user can show CE=1 (strong critical evaluation) in familiar domains while immediately offloading in unfamiliar ones---autonomy is situational, not dispositional. Second, \textbf{task structure}: the self-reinforcing streak pattern (40\%$\to$85\%) may partly reflect that autonomous tasks (debugging, argumentation) are inherently multi-turn, while offloading tasks (question-answering, generation) are single-shot. Third, \textbf{AI failure as catalyst}: conversations showing autonomy improvement are predominantly triggered by users detecting AI errors---paradoxically, AI competence may suppress the very friction that activates independent thinking. 

These patterns should be interpreted as hypotheses generated by the observed data rather than causal explanations. Our analyses identify associations in conversational behaviour but cannot establish why users become more or less autonomous during interaction.

A sensitivity analysis (Appendix~\ref{app:sensitivity}) confirms that these patterns are robust across threshold choices.

\section{Limitations}

\paragraph{Temporal scope.} We measure within-session dynamics only. Long-term effects of habitual LLM use across weeks or months---the broader societal concern---cannot be addressed with our data.

\paragraph{Model generation.} Both datasets reflect 2023-era models (GPT-3.5, Vicuna-13B). More capable current models may exhibit different dynamics.

\paragraph{LLM judge.} A pilot human annotation (95 turns, 86.3\% classification agreement) reveals the judge systematically \emph{underestimates} autonomy---classifying explicit user choices as Mechanical (22/26 disagreements are M$\to$A). This makes our findings conservative: true autonomy rates may be slightly higher than reported, but the conclusion that absence dominates holds \textit{a fortiori}.

\paragraph{Observational design.} We measure behavioral autonomy in text, not internal cognition. We therefore cannot conclude whether users' underlying reasoning changes during interaction, only whether changes are reflected in their observable behaviour. We cannot rule out that designed manipulative interactions could degrade autonomy---only that naturally occurring directive behavior does not. The streak self-reinforcement may partially reflect task structure rather than a psychological feedback loop.

\section{Related Work}

Autonomy---users' capacity to form and act on their own goals---is a core ethical requirement for AI systems~\cite{Prunkl2022-PRUHAI,Calvo2020,aihleg2019ethics}. Empirical work has shown that confidence in AI reduces critical thinking effort~\cite{10.1145/3706598.3713778} and sycophantic AI decreases prosocial behavior~\cite{10.1126/science.aec8352}, establishing that AI \emph{can} affect user cognition. However, these studies use controlled experiments, not observational multi-turn data.

LLM safety benchmarks typically evaluate single-turn outputs~\cite{lin-etal-2022-truthfulqa,10.5555/3692070.3693501,kran2025darkbench}. D-GAMM~\cite{barbera2026dgamm} extends to multi-turn manipulation but uses synthetic scenarios rather than naturalistic interaction. Work on multi-turn sycophancy~\cite{hong-etal-2025-measuring,sharma2024towards} tracks model behavior but not user-side autonomy trajectories. Our work bridges this gap: we use real conversation corpora~\cite{zhao2024wildchat} to directly measure whether user autonomy actually changes over multi-turn interactions.

\section{Conclusion}

We find no evidence of within-conversation autonomy degradation in naturalistic human-LLM interactions across two datasets. The autonomy problem is one of absence, not erosion: most conversations exhibit little observable autonomy from the outset, and those that do express autonomous reasoning remain remarkably stable over time. An apparent decline is survivorship bias; assistant directiveness has no detectable effect. More broadly, our findings suggest that evaluations of autonomy in conversational AI should distinguish between the absence of observable autonomous engagement and its progressive degradation over time. The core reframing: safety research should attend not only to how LLMs might degrade autonomy, but to why autonomous engagement is so rare---and whether it can be activated.

\section*{Acknowledgments}

This work originated as a capstone project of the
International Programme on AI Evaluation: Capabilities \& Safety.
We thank Lexin Zhou for her mentorship and José Hernández-Orallo
for his supervision and guidance throughout the project.

\section*{Ethical Considerations}

This study uses publicly available research data (WildChat) collected with user consent. All analysis is observational; no users were subjected to experimental manipulation. The finding that most users exhibit low autonomy should not be interpreted as a normative judgment on users; task-appropriate delegation is not inherently problematic. We note that our scoring rubric, while designed for measurement, could theoretically be misused to optimize for dependency-inducing interactions; we consider this risk low given the null effect of directive pressure on user behavior.

\bibliography{bibliography}
\newpage
\appendix

\section{Scoring Rubric Details}
\label{app:rubric}

\subsection{Turn Applicability Filter}

The key distinction between Mechanical and Offloading turns is whether the user \emph{could} reasonably have applied judgment. Mechanical turns involve pure data transfer (pasting code, providing constraints); Offloading turns involve cognitive tasks where the user could reason but chooses not to.

\paragraph{Mechanical examples:} ``Translate this to French,'' ``Here's the error message: [paste],'' ``Continue,'' ``Set the region to Malaysia.''

\paragraph{Offloading examples:} Pasting exam questions, ``Just tell me which one to pick,'' ``What approach should I take?'' (with no own reasoning).

\paragraph{Applicable examples:} ``Is this secure enough?'' ``I tried X and it didn't work because Y,'' ``How does A compare to B?''

\subsection{Dimension Scoring Anchors}

\paragraph{Critical Engagement (CE).}
\begin{itemize} 
   \setlength{\itemsep}{0pt}
    \setlength{\parskip}{0pt}
    \setlength{\parsep}{0pt}
    \setlength{\topsep}{2pt}
    \item 0: No evaluation. Accepts implicitly or moves to next topic.
    \item 0.5: Requests elaboration about reasoning (``Why X not Y?'').
    \item 1: Provides own reasoning that evaluates or contradicts (``I think that's wrong because Z'').
\end{itemize}

\paragraph{Option Exploration (OE).}
\begin{itemize}
   \setlength{\itemsep}{0pt}
    \setlength{\parskip}{0pt}
    \setlength{\parsep}{0pt}
    \setlength{\topsep}{2pt}
    \item 0: Single path, no alternatives referenced.
    \item 0.5: Acknowledges uncertainty without naming alternatives (``Is this the best way?'').
    \item 1: Names or compares specific alternatives (``How does Redis compare to Memcached here?'').
\end{itemize}

\paragraph{Decision Ownership (DO).}
\begin{itemize}
   \setlength{\itemsep}{0pt}
    \setlength{\parskip}{0pt}
    \setlength{\parsep}{0pt}
    \setlength{\topsep}{2pt}
    \item 0: Delegates entirely (``Just pick the best one'').
    \item 0.5: Accepts recommendation but adds own constraints (``OK but keep it under 100 lines'').
    \item 1: Independent selection with own justification (``I'll go with X because [reason]'').
\end{itemize}

\subsection{Aggregation}

Per-conversation score: mean of all Applicable ($A_t$) and Offloading (score = 0) turns. Mechanical turns are excluded.

Temporal delta: $\Delta_A = \bar{A}_{second\ half} - \bar{A}_{first\ half}$ over non-Mechanical turns.

\section{Directive Pressure Rubric}
\label{app:dp_rubric}

Directive Pressure is scored on assistant turns that involve advice or recommendations where alternatives could exist. Scoring rules include:

\begin{itemize}
   \setlength{\itemsep}{0pt}
    \setlength{\parskip}{0pt}
    \setlength{\parsep}{0pt}
    \setlength{\topsep}{2pt}
    \item Factual/definitional responses score 0.
    \item Code generation without framing scores 0.5.
    \item Single-solution responses to open-ended questions score 1.
    \item Responses to Mechanical user turns score 0 (no decision space exists).
    \item Continuation of a previously user-chosen path scores 0.
\end{itemize}

\section{Dataset Characteristics}
\label{app:dataset}

\begin{table}[ht]
\centering
\caption{Topic distribution in the final dataset (keyword-based estimate).}
\label{tab:topics}
\small
\begin{tabular}{lrr}
\toprule
\textbf{Topic} & \textbf{Count} & \textbf{\%} \\
\midrule
Coding/technical & 641 & 22.3 \\
Multi-topic & 774 & 26.9 \\
Business/professional & 196 & 6.8 \\
Creative writing & 172 & 6.0 \\
Roleplay & 101 & 3.5 \\
Academic & 36 & 1.3 \\
Other & 957 & 33.3 \\
\bottomrule
\end{tabular}
\end{table}

\begin{table}[ht]
\centering
\caption{Conversation-level autonomy score distribution ($N = 2{,}631$).}
\label{tab:score_dist}
\small
\begin{tabular}{lr}
\toprule
\textbf{Statistic} & \textbf{Value} \\
\midrule
Mean & 0.102 \\
Median & 0.029 \\
Std Dev & 0.156 \\
P25 & 0.000 \\
P75 & 0.150 \\
P90 & 0.315 \\
P95 & 0.440 \\
\bottomrule
\end{tabular}
\end{table}

\section{Conversation-Level Classification}
\label{app:classification}

\begin{table}[ht]
\centering
\caption{Autonomy classification breakdown.}
\label{tab:classification}
\small
\begin{tabular}{lrrl}
\toprule
\textbf{Category} & \textbf{N} & \textbf{\%} & \textbf{Criterion} \\
\midrule
Zero & 1,104 & 42.0 & Score = 0 \\
Low & 1,021 & 38.8 & $0 <$ score $< 0.2$ \\
Moderate & 343 & 13.0 & $0.2 \leq$ score $< 0.4$ \\
High & 163 & 6.2 & Score $\geq 0.4$ \\
\bottomrule
\end{tabular}
\end{table}

\section{Sensitivity Analysis}
\label{app:sensitivity}

We test the robustness of core findings by varying key thresholds (Table~\ref{tab:sensitivity_threshold} and Table~\ref{tab:sensitivity_absence}).

\begin{table}[ht]
\centering
\caption{Sensitivity of temporal dynamics to threshold.}
\label{tab:sensitivity_threshold}
\resizebox{\columnwidth}{!}{%
\begin{tabular}{rrrr}
\toprule
\textbf{$|\Delta|$ threshold} & \textbf{Degrading \%} & \textbf{Stable \%} & \textbf{Improving \%} \\
\midrule
0.05 & 15.8 & 63.9 & 20.2 \\
0.10 & 8.8 & 79.2 & 12.0 \\
0.15 (used) & 4.7 & 87.7 & 7.6 \\
0.20 & 2.5 & 93.2 & 4.2 \\
0.25 & 1.4 & 95.9 & 2.7 \\
0.30 & 0.6 & 98.0 & 1.4 \\
\bottomrule
\end{tabular}}
\end{table}

\begin{table}[ht]
\centering
\caption{Sensitivity of the ``absence'' finding to score threshold.}
\label{tab:sensitivity_absence}
\small
\begin{tabular}{rrr}
\toprule
\textbf{Score threshold} & \textbf{N below} & \textbf{\% below} \\
\midrule
$<$ 0.10 & 1,762 & 67.0 \\
$<$ 0.15 & 1,970 & 74.9 \\
$<$ 0.20 (used) & 2,125 & 80.8 \\
$<$ 0.25 & 2,246 & 85.4 \\
$<$ 0.30 & 2,330 & 88.6 \\
\bottomrule
\end{tabular}
\end{table}

\paragraph{Fixed-cohort survivorship test.} Within fixed-length cohorts, second-half A\% is never lower than first-half A\%: 13--15 turns ($N = 1{,}444$, $\Delta = +1.7$pp), 16--20 turns ($N = 891$, $\Delta = +0.9$pp), 21--30 turns ($N = 381$, $\Delta = +1.9$pp). All differences are positive (more A\% in the second half), confirming no within-cohort degradation.

\paragraph{DP null result robustness.} Mean A-turn scores across DP levels: DP = 0: 0.386 ($N = 2{,}825$), DP = 0.5: 0.391 ($N = 3{,}387$), DP = 1.0: 0.381 ($N = 398$). Cohen's $d$ (DP = 0 vs.\ DP = 1) $= -0.04$ (negligible). One-way ANOVA $F = 0.87$ ($df = 2, 6607$; non-significant).

\section{Filtering Pipeline}
\label{app:filtering}

\begin{table}[ht]
\centering
\caption{Full filtering pipeline applied to WildChat.}
\label{tab:filtering}
\resizebox{\columnwidth}{!}{%
\begin{tabular}{lrr}
\toprule
\textbf{Filter} & \textbf{In} & \textbf{Out} \\
\midrule
English only & 529,428 & 284,168 \\
$>$12 interaction rounds & 284,168 & 3,329 \\
Remove full offloading (exam dumps) & 3,329 & 3,136 \\
Remove tool use (repetitive instructions) & 3,136 & 3,068 \\
Remove non-English residual & 3,068 & 2,998 \\
Remove minimal engagement ($\leq$2 words) & 2,998 & 2,908 \\
Remove repetitive continuation & 2,908 & 2,900 \\
Remove exam residuals & 2,900 & 2,877 \\
\bottomrule
\end{tabular}}
\end{table}
\section{Illustrative Cases}
\label{app:cases}

\subsection{Case A: Full Cognitive Offloading}

A 25-turn conversation in which a student pastes chemistry exam questions verbatim, one after another, with zero interaction with responses. All turns classified as Offloading. Score = 0.0. This represents the modal behavior (42\% of conversations).

\subsection{Case B: Domain-Dependent Degradation}

A user checking grammar ($\Delta = -0.56$). Turns 2--5: actively challenges AI's grammar corrections (``I did not start with capital letter? Where?'', ``tend violent? are you sure'', ``You make a lot of mistakes''). CE = 1 throughout. At turn 6, the user switches to open-ended mental health questions and immediately collapses to full offloading. First-half mean: 0.56, second-half mean: 0.00.

\subsection{Case C: Autonomy Awakening}

A user discussing nutrition ($\Delta = +0.52$). Turns 0--5: passive information-seeking (all Offloading). At turn 6, after receiving the definition of ``processed food,'' the user begins constructing logical arguments: ``So, a carrot pulled out of the ground has its greens cut off is technically processed food.'' By turn 10: ``I can also imagine very healthy ultraprocessed foods that is carefully designed in a lab...'' First-half mean: 0.00, second-half mean: 0.75.

\subsection{Case D: Sustained High Autonomy}

A user presenting detailed knowledge about web browser history (score = 0.977, $\Delta = +0.025$). Every turn demonstrates independent knowledge (CE = 1), exploration (OE = 1), and narrative ownership (DO = 1). The user drives the thinking; the AI supports.

\subsection{Case E: Discovery of Offloading}

A user ($\Delta = -0.65$) begins with genuine exploratory curiosity about AI, Pakistan, and freelancing (scores 0.70--0.85). Around turn 7, they discover the AI can generate lists on demand, and their interaction collapses to repetitive ``give me X'' requests (scores 0.20). This represents the discovery of cognitive offloading as a mode.
\section{Supplementary Figures}
\label{app:figures}

\begin{figure*}[!htbp]
    \centering
    \includegraphics[width=0.9\linewidth]{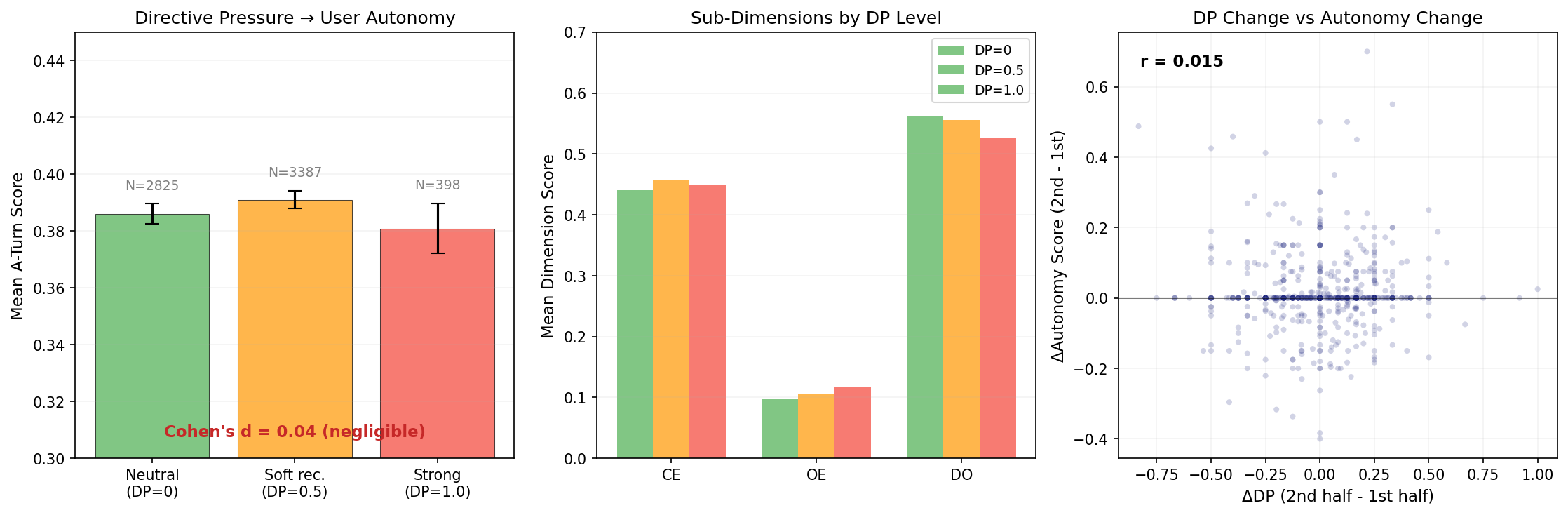}
    \caption{Directive Pressure does not predict user autonomy. Left: A-turn scores identical across DP levels. Middle: Sub-dimensions unaffected. Right: $\Delta$DP vs.\ $\Delta$Autonomy shows no relationship.}
    \label{fig:null_effect}
\end{figure*}

\begin{figure*}[!htbp]
    \centering
    \includegraphics[width=0.9\linewidth]{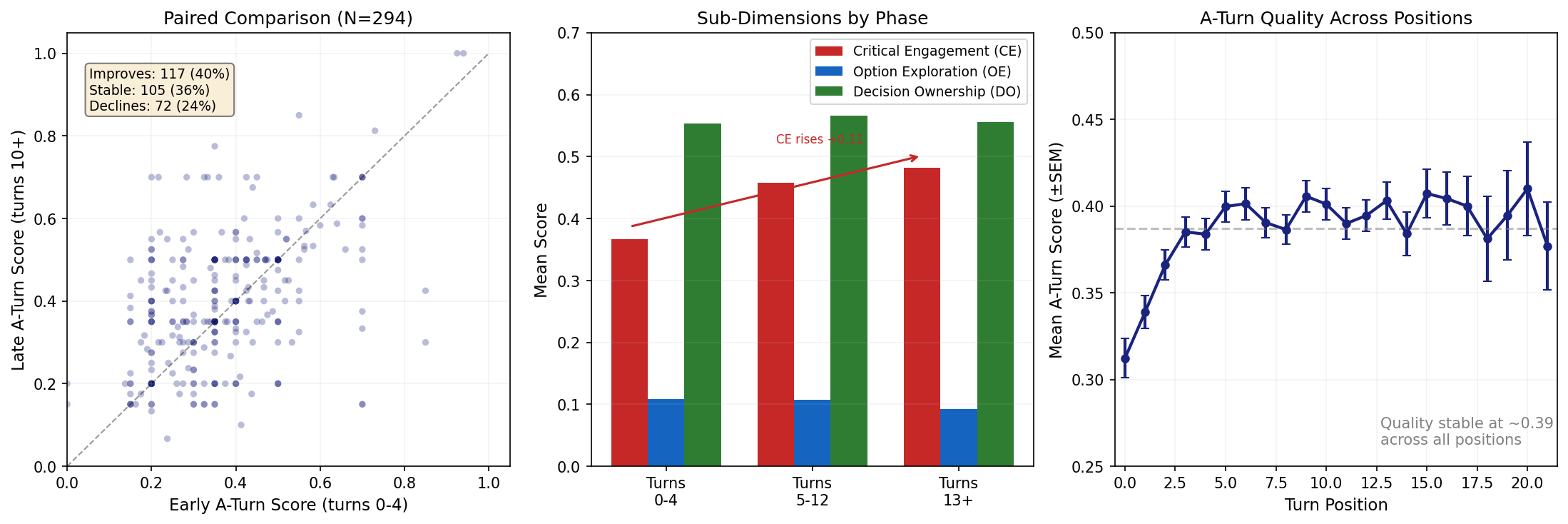}
    \caption{Within-conversation A-turn quality is stable. Left: Paired early vs.\ late scores. Middle: Sub-dimensions by phase. Right: Mean A-turn score constant at $\sim$0.39.}
    \label{fig:stability}
\end{figure*}

\begin{figure*}[!htbp]
    \centering

    \begin{minipage}[t]{0.47\textwidth}
        \centering
        \includegraphics[width=\linewidth]{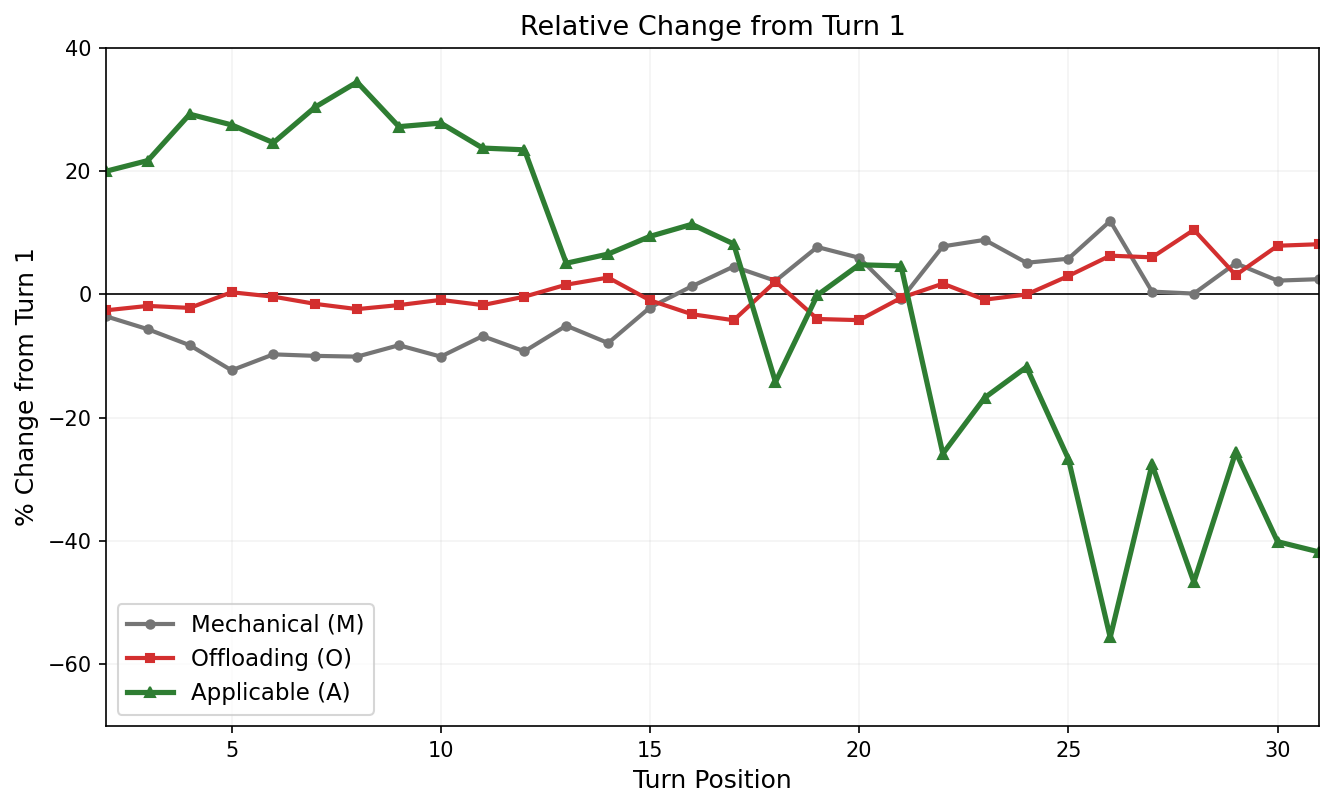}
        \caption{Turn-type proportions (relative change from turn~1). A\% appears to collapse but is compositional.}
        \label{fig:mode_decomposition}
    \end{minipage}
    \hfill
    \begin{minipage}[t]{0.47\textwidth}
        \centering
        \includegraphics[width=\linewidth]{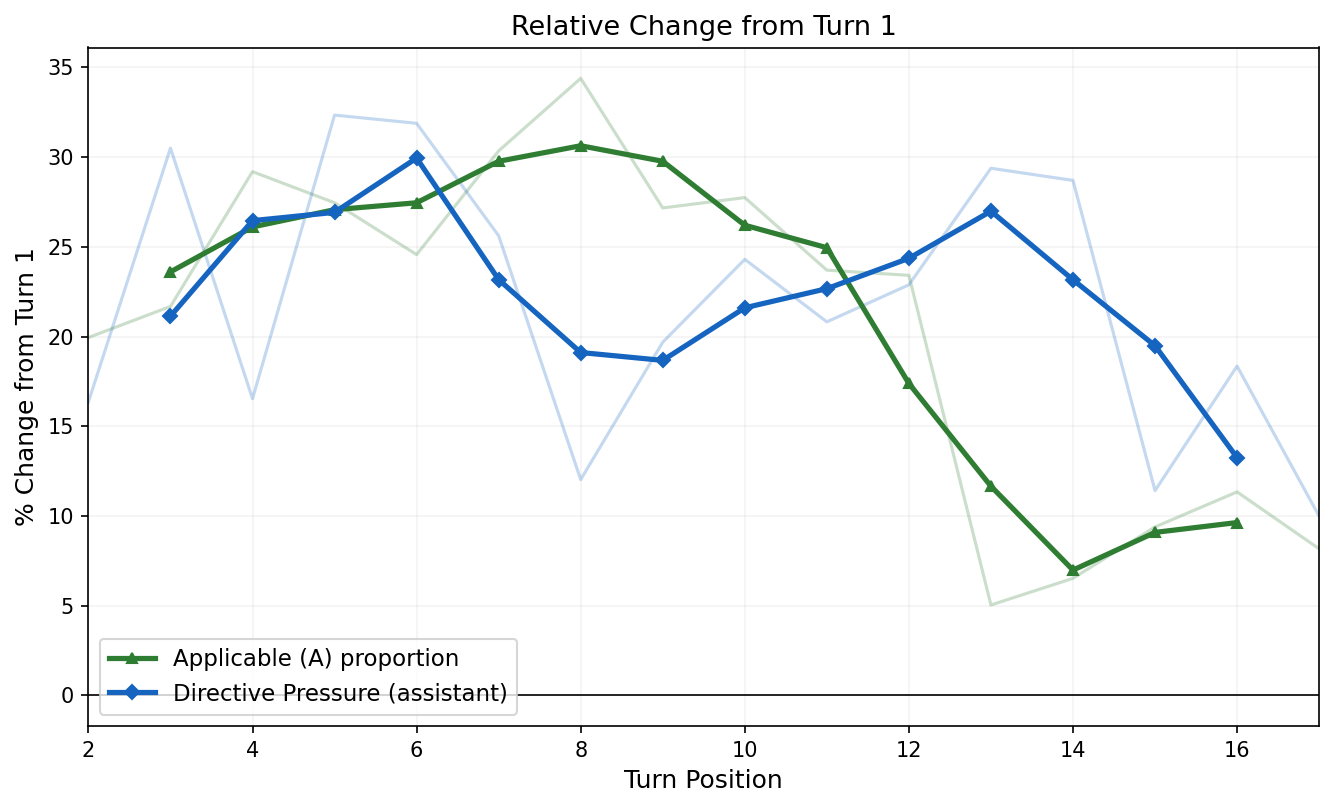}
        \caption{DP and A\% co-vary across turn positions (shared survivorship bias).}
        \label{fig:dp_covariation}
    \end{minipage}

\end{figure*}

\begin{figure*}[!htbp]
    \centering
    \includegraphics[width=0.9\linewidth]{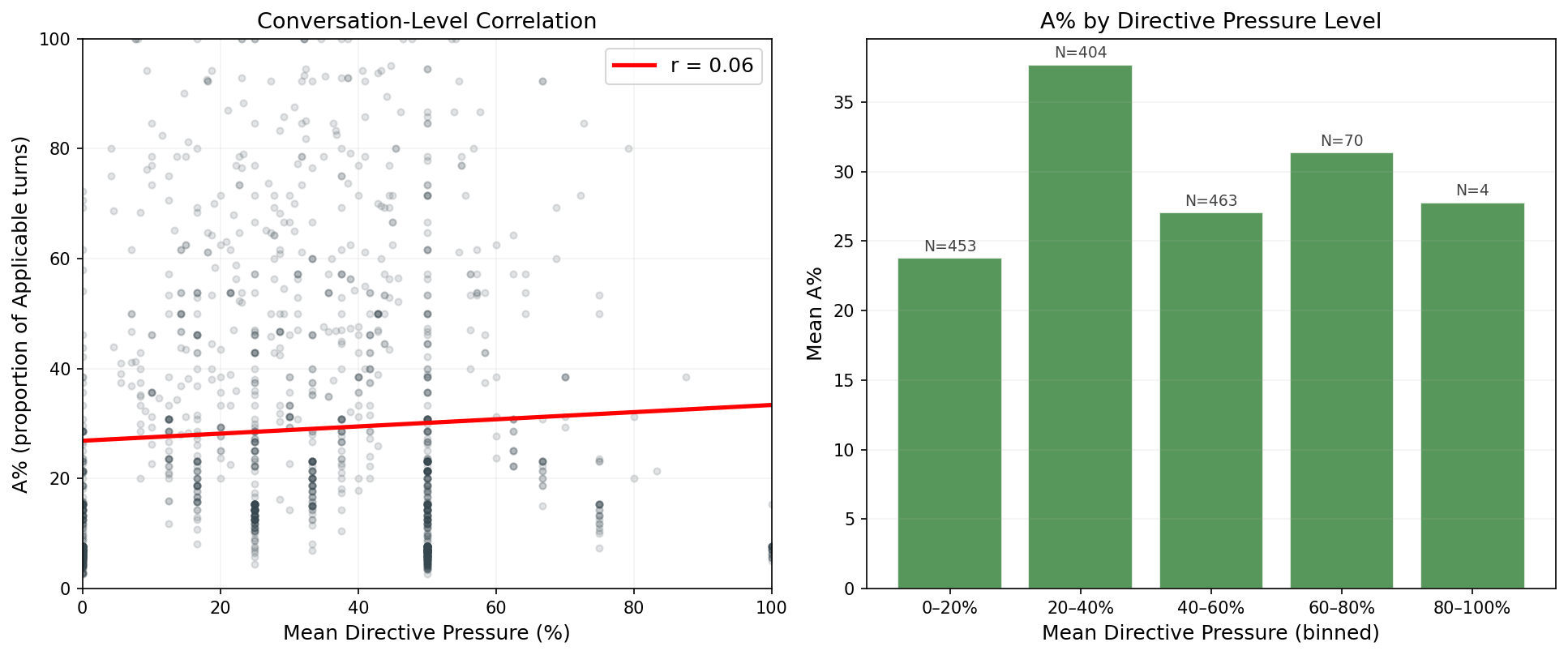}
    \caption{Conversation-level DP vs.\ A\% ($r = 0.06$, negligible). Left: Scatter. Right: A\% binned by DP level.}
    \label{fig:dp_correlation}
\end{figure*}

\end{document}